\documentclass[a4paper]{article}
\usepackage[numbers,sort&compress]{natbib}
\usepackage{INTERSPEECH2020}
\usepackage{multirow}
\usepackage{subfigure}
\usepackage{color}

\title{Spoofing-Aware Speaker Verification by Multi-Level Fusion}
\name{Haibin Wu$^{1,2}$, Lingwei Meng$^{3}$, Jiawen Kang$^{3}$, Jinchao Li$^{3}$, Xu Li$^{3}$, Xixin Wu$^{3}$, \\ Hung-yi Lee$^{1}$, Helen Meng$^{2,3}$
}

\address{
  $^1$ Graduate Institute of Communication Engineering, National Taiwan University\\
  $^2$ Centre for Perceptual and Interactive Intelligence, The Chinese University of Hong Kong \\
  $^3$ Human-Computer Communications Laboratory, The Chinese University of Hong Kong 
  }

\begin{document}

\maketitle
\begin{abstract}
Recently, many novel techniques have been introduced to deal with spoofing attacks, and achieve promising countermeasure (CM) performances.
However, these works only take the stand-alone CM models into account.
Nowadays, a spoofing aware speaker verification (SASV) challenge which aims to facilitate the research of integrated CM and ASV models, arguing that jointly optimizing CM and ASV models will lead to better performance, is taking place.
In this paper, we propose a novel multi-model and multi-level fusion strategy to tackle the SASV task.
Compared with purely scoring fusion and embedding fusion methods, this framework first utilizes embeddings from CM models, propagating CM embeddings into a CM block to obtain a CM score.
In the second-level fusion, the CM score and ASV scores directly from ASV systems will be concatenated into a prediction block for the final decision.
As a result, the best single fusion system has achieved the SASV-EER of 0.97\% on the evaluation set.
Then by ensembling the top-5 fusion systems, the final SASV-EER reached 0.89\%.
\end{abstract}


\section{Introduction}
\label{sec:intro}
Spoofing attacks and countermeasures for automatic speaker verification (ASV) have aroused keen interests in academia and the industry. 
While ASV systems aim to verify the identity of target speakers, spoofing attacks attempt to manipulate the verification results using various technologies, leading to dramatic performance degradation \cite{wu2015asvspoof, kinnunen2017asvspoof, todisco2019asvspoof, yamagishi2021asvspoof, wang2021comparative}.
In order to ensure the robustness and security of ASV systems, CM is a necessary technique to detect and defend spoofing attacks.

The vulnerability of ASV systems was revealed in \cite{wu2015asvspoof, kinnunen2017asvspoof, todisco2019asvspoof, yamagishi2021asvspoof}, under speech synthesis and voice conversion (VC) attacks. 
Currently, various techniques have been proposed to perform effective attacks, including audio replay \cite{wu2014study}, adversarial noise \cite{kreuk2018fooling, das2020attacker}, more advanced text-to-speech (TTS) and VC models \cite{de2010evaluation, wu2014voice, kons2013voice}.
In addition, many works have been done to investigate state-of-the-art CM strategies.
The current solutions leverage end-to-end deep neural networks (DNNs) \cite{monteiro2019development, monteiro2020generalized}, trying to distinguish artifacts and unnatural cues of spoofing speech from bona fide speech.
And thanks to a series of challenges and datasets \cite{wu2015asvspoof, kinnunen2017asvspoof, todisco2019asvspoof, yamagishi2021asvspoof}, many novel techniques were introduced to achieve promising CM performances \cite{monteiro2019development, monteiro2020generalized,zhang2021one, wu2022partial,chen2020generalization, jung2021aasist,li2021replay, tak2021graph, jung2019replay,shim2020integrated,jung2020study,shim2019self, tak2022automatic,tak2021end,kamble2018effectiveness,tak2020spoofing,tak2018novel}. 

However, previous works only take the stand-alone CM models into account.
Recently, a spoofing aware speaker verification (SASV) challenge \cite{jung2022sasv} was proposed as a special session in ISCA INTERSPEECH 2022.
This challenge aims to facilitate the research of integrated CM and ASV models, arguing that jointly optimizing CM and ASV models will lead to better performance.
To measure the performance of integrated models, a SASV-EER was proposed in this challenge as a primary metric, which is a variant of classic equal error rate (EER). Under this metric, the test utterances in trials belong to one of three types: impostors, target speakers, and spoofing attacks.
In further, the SASV-EER can be subsetted into SV-EER (impostors vs. targets) and SPF-EER (targets vs. spoof). The former is for evaluating speaker verification performance, and the latter is for evaluating anti-spoofing performance.
In this way, this metric expects the model can accept target speakers and reject any alternatives, including the impostors and spoofing attacks.
This metric is a straightforward assessment for integrated SASV systems.

This paper described our submitted system for the SASV Challenge 2022.
In order to take advantage of existing well-designed models in CM and ASV areas, we proposed a novel multi-model and multi-scale fusion framework.
Compared with purely scoring fusion and embedding fusion methods, this framework utilizes embeddings from CM models in the first-level fusion, propagating CM embeddings into a CM block to obtain a CM score.
In the second-level fusion, the CM score and ASV scores directly from ASV systems will be concatenated into a prediction block for the final decision.
In contrast to our previous work \cite{wu2022sasv} that only simply concatenates the embeddings from different CM models, we considered the potentials of pooling strategies in terms of feature aggregation, and investigated various pooling methods \cite{you2019multi, snyder2018x, zhu2018self, okabe2018attentive} when fusing embeddings across different CM models.
Based on the proposed fusion framework, we presented the fusion strategies of a series of state-of-the-art CM and ASV models with different pooling strategies to boost the fusion results.
As a result, the best single fusion system has achieved the SASV-EER of 0.97\% on the evaluation set.
Then by ensembling the top-5 fusion systems, the final SASV-EER reached 0.89\% on the evaluation set, while this number in the best baseline system from the SASV challenge is 6.37\%.

\begin{figure*}[ht]
  \centering
  \centerline{\includegraphics[width=1.0\linewidth]{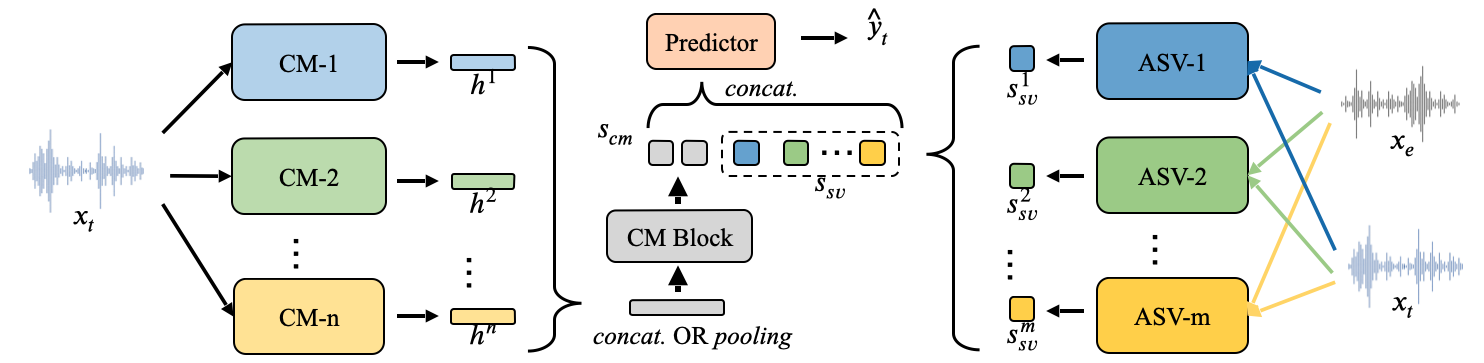}}
  \caption{The proposed multi-model \& multi-level fusion framework.}
  \label{fig:method}
\end{figure*}
\section{Background}

\subsection{SASV strategies}
Given the enrollment utterance $x_{e}$ and the testing utterance $x_{t}$, spoofing-aware speaker verification (SASV) systems aim at telling $y_{t} = 1$ if $x_{t}$ comes from the same speaker as $x_{e}$, or $y_{t} = 0$ if $x_{t}$ comes from another speaker or $x_{t}$ is a spoofing attack. There are two typical strategies for constructing a SASV system: multi-task learning strategy and fusion-based strategy. 

The multi-task learning strategy trains the models jointly with both speaker verification and anti-spoofing objectives, which is intuitive to be adopted. 
The two objectives share the same backbone and thus the features and embeddings, while each objective has its own predicting head and loss function. 
It is worth noting that speaker verification and anti-spoofing objectives are contradictory in some respects. 
The former drives the model to erase device and environment information to more robustly identify speakers; in contrast, the latter prompts the model to capture device and environment traces, then tells forged spoofing from authentic utterances \cite{shim2020integrated}.

Alternatively, the fusion-based strategy has the potential to reach better SASV performance leveraging state-of-the-art CM and ASV models trained on large-scale datasets. 
Considering this superiority, we propose our solutions for the SASV challenge based on a novel multi-model and multi-level fusion strategy. 

\subsection{Baseline systems}
The challenge organizer provides two baseline systems. Each system is based upon ECAPA-TDNN model \cite{desplanques2020ecapa} as the ASV subsystem and AASIST model \cite{jung2021aasist} as the CM subsystem. The key difference between the two systems is how they fuse ASV and CM subsystem - Baseline1 adopts the score-level fusion, while Baseline2 adopts the embedding-level fusion. 

\section{Method}

\begin{figure*}[ht]
\centering
\subfigure[ECAPA-TDNN]{
\centering
\includegraphics[width=0.32\linewidth]{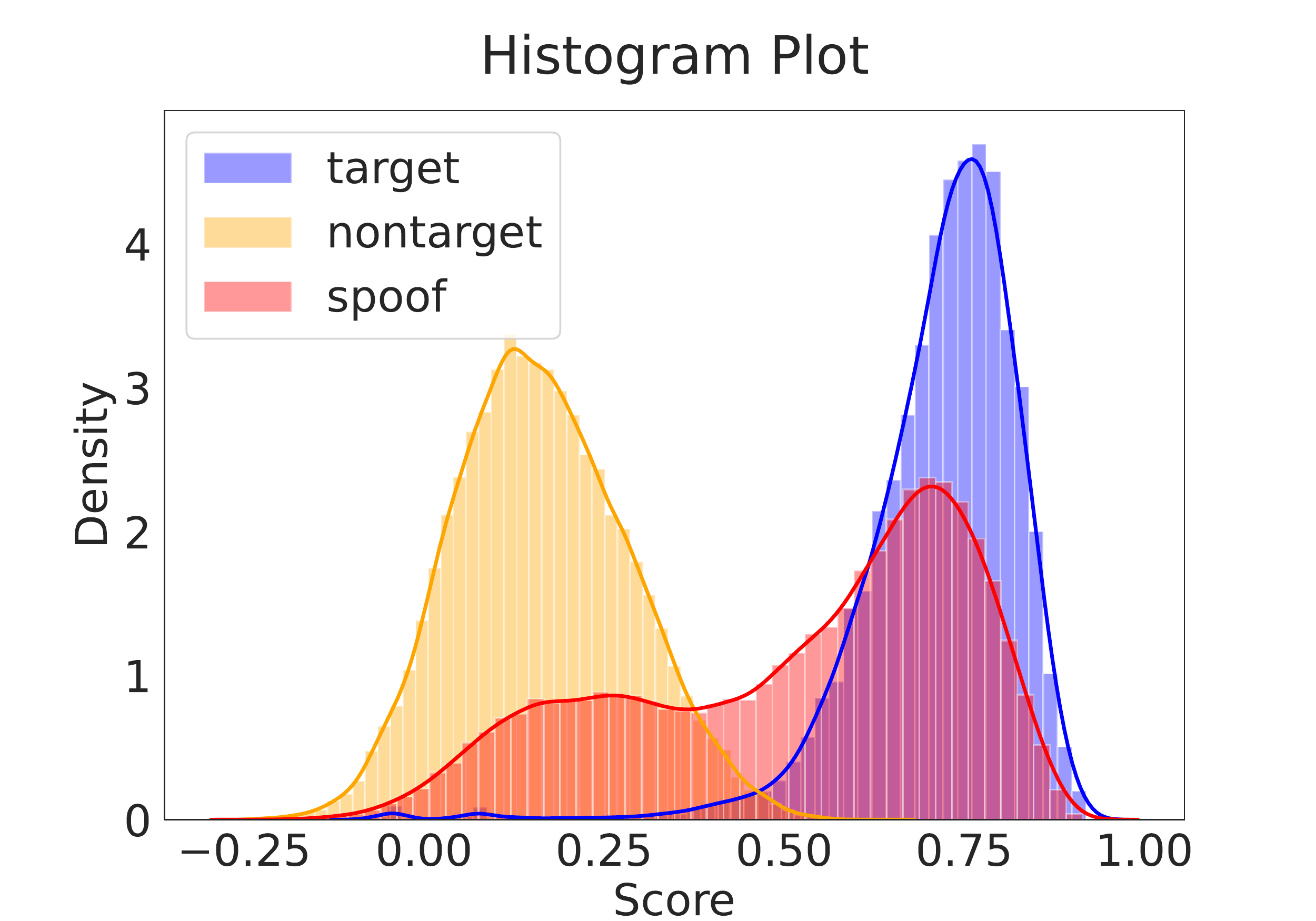}
}
\subfigure[AASIST]{
\centering
\includegraphics[width=0.32\linewidth]{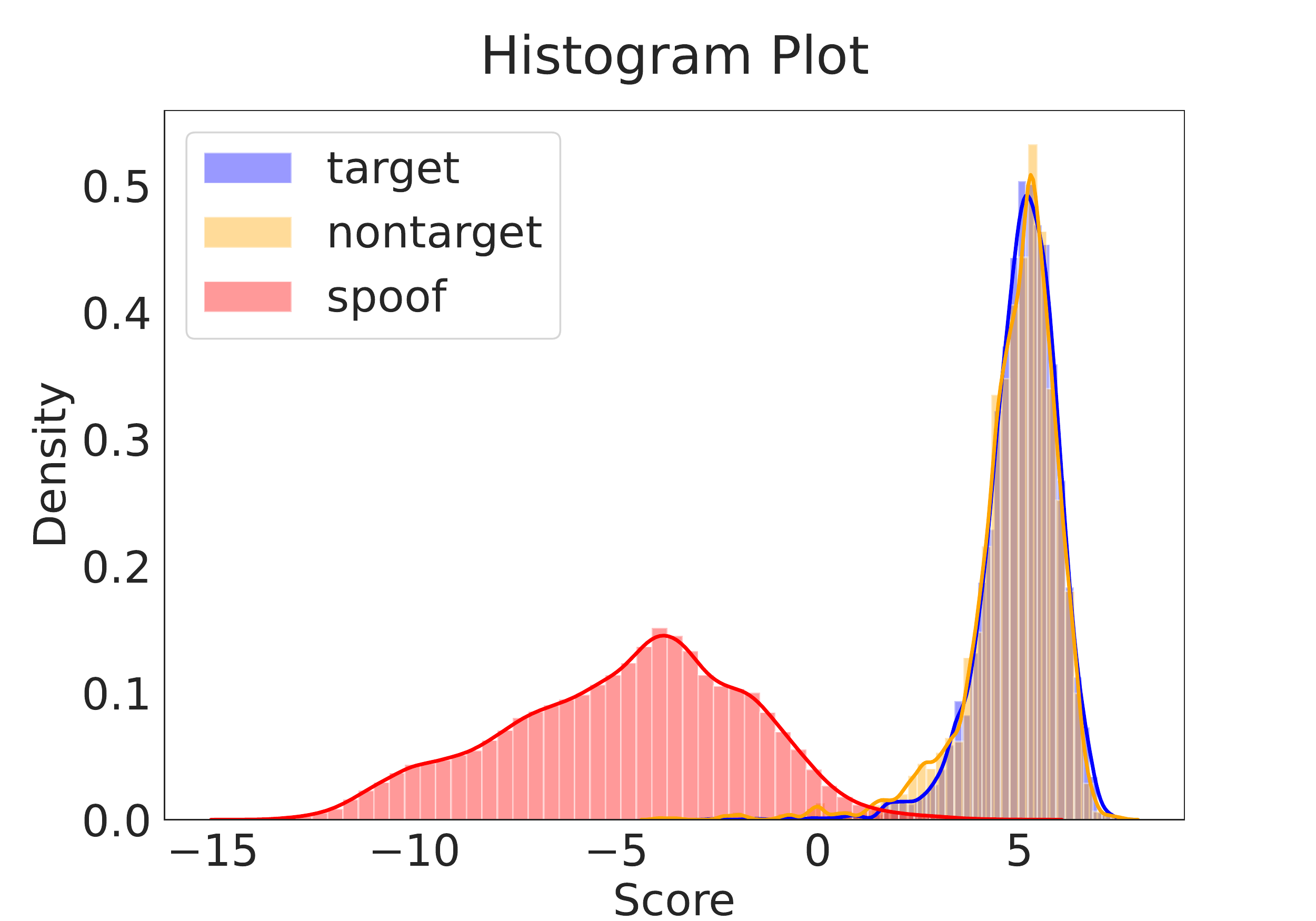}
}
\subfigure[Proposed top-5 ensemble system]{
\centering
\includegraphics[width=0.32\linewidth]{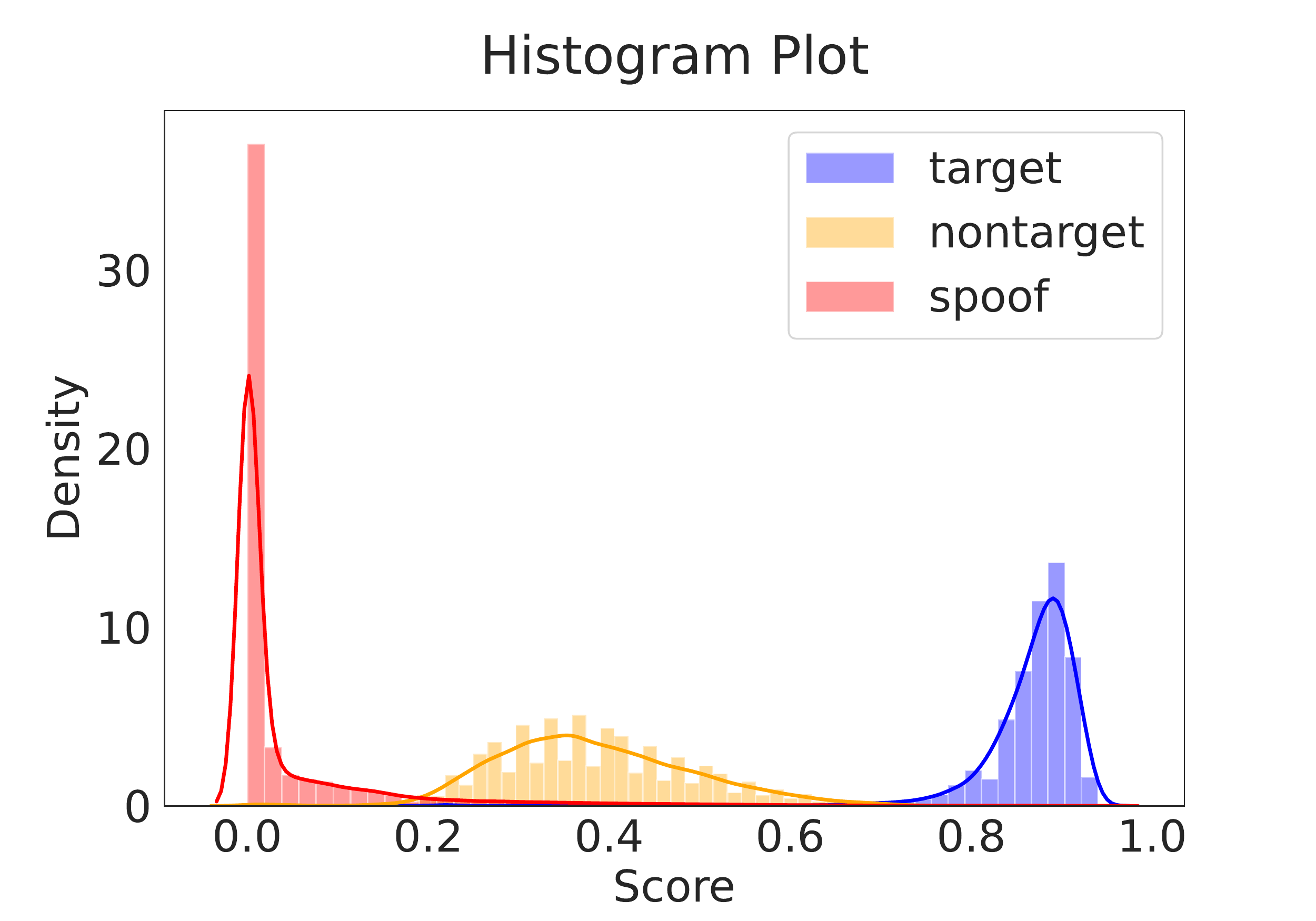}
}
\centering
\caption{The histogram plots of the output scores predicted by ASV, CM, and the proposed top-5 ensemble system. Other proposed system variants also have similar histogram shapes as (c).}
\label{fig:hist}
\end{figure*}

Although achieving acceptable performances, baseline systems' fusion strategies are relatively simple and naive. Baseline1's score-level fusion does not guarantee that the ASV score and the CM score belong to a unified space and an identical magnitude. Baseline2 crudely concatenates three embeddings and throws the product into a DNN, lacking fine-grained fusion. These inadequacies motivate us to explore further possibilities of fusion strategies in the SASV task.

In this paper, we proposed a multi-model and multi-level fusion strategy. In terms of the width, it employs multiple pre-trained ASV and CM models as plug-and-play components, where users can expand or shrink the scale of the model according to their needs. In terms of the depth of this strategy, it fuses CM embeddings to calculate a score in the first-level fusion, which is then integrated with the ASV models' outputs, and yields the final prediction in the second-level fusion.

\subsection{Overall structure}
The overall framework is shown in Figure \ref{fig:method}, where $x_{e}$ and $x_{t}$ are the input enrollment and testing utterances, respectively; \textit{ASV-1}, \textit{ASV-2}, ..., \textit{ASV-m} denote $m$ pre-trained ASV models; \textit{CM-1}, \textit{CM-2}, ..., \textit{CM-n} denote $n$ pre-trained CM models. Given a trial $\{x_{e}, x_{t}\}$, a series of cosine scores $\{s_{sv}^1, s_{sv}^2, \ldots, s_{sv}^m\}$ are derived from $m$ ASV models. Given the testing utterance $x_{t}$, a series of CM embeddings $\{h^{1}, h^{1}, \ldots, h^{n}\}$ are extracted through $n$ CM models. Next comes the first level of fusion, where the $n$ embeddings are integrated into a $h_{cm}$ by concatenation or a pooling method.
Our previous work \cite{wu2022sasv} investigated the capacity of concatenation for SASV. 
In this work, we further extend the potential of concatenation by making the CM block in Figure \ref{fig:method} deeper, and we also adopt pooling methods to further improve the fusion performance.
Further, $h_{cm}$ goes through a CM Block to better digest fused embeddings and then a 2-dimension countermeasure score $s_{cm}$ is predicted. With $s_{cm}$ and $\{s_{sv}^1, s_{sv}^2, \ldots, s_{sv}^m\}$ well prepared, they are concatenated and fed into the Predictor to yield the final prediction $\hat{y_t}$, where the second-level fusion is performed.

\subsection{Strategies in the first-level fusion}
For the first-level fusion, we attempt concatenation \cite{wu2022sasv} or one of the four kinds of pooling methods to synthesize $h_{cm}$ separately, and have conducted extensive experiments accordingly.
Suppose $\bm{H}=\{h^{1}, h^{1}, \ldots, h^{n}\}$, and the length of all the CM embeddings are projected into a same length $d_h$ by feed forward layers before performing concatenation or pooling. The four candidate pooling methods are Temporal Average Pooling (TAP) \cite{you2019multi}, Temporal Statistics Pooling (TSP) \cite{snyder2018x}, Self-attentive Pooling (SAP) \cite{zhu2018self}, and Attentive Statistics Pooling (ASP) \cite{okabe2018attentive}.

TAP is to calculate the mean value along the channels to obtain the $h_{cm}$. 

TSP calculates channel-wise mean and standard deviation, then concatenates the mean vector and standard deviation vector together as $h_{cm}$.

In SAP, the self-attention mechanism takes $\bm{H}$ as input and outputs an annotation matrix $\bm{A}$:
\begin{equation}
     \bm{A} = softmax(tanh(\bm{H}^T\bm{W}_1)\bm{W}_2)
    \label{eq:SAP-weight}
\end{equation}
where $\bm{W}_1$ is a matrix of size $d_h\times d_a$; $\bm{W}_2$ is a matrix of size $d_a\times d_r$, and $d_r$ is a hyper-parameter that represents the number of attention heads; The $softmax(·)$ is performed column-wise. We set $d_r=1$, therefore $\bm{A}$ degenerates into an annotation vector. Weighted by $\bm{A}$, $h_{cm}$ is calculated as the weighted mean:
\begin{equation}
     h_{cm} =\Tilde{\mu} = \bm{HA}
    \label{eq:SAP-mean}
\end{equation}

For ASP, not only it calculate a attention-weighted mean as SAP do, but also it calculate a attention-weighted standard deviation:
\begin{equation}
     \Tilde{\sigma} =\sqrt{\sum_{i=1}^{n}{\alpha^i h^i\odot h^i }-\Tilde{\mu}\odot\Tilde{\mu}}
    \label{eq:ASP-sigma}
\end{equation}
where $\alpha^i$denotes the $i^{th}$ element of the annotation vector $\bm{A}$, $\odot$ represents the Hadamard product. By concatenating  $\Tilde{\mu}$ and $\Tilde{\sigma}$, $h_{cm}$ is derived.

After one of the above pooling methods or concatenation, derived $h_{cm}$ goes through CM Block, which is a multi-layer perceptron, and generates a two-dimension score reflecting the possibilities the testing utterances is the target or not.

\subsection{Loss function}
The ASV and CM models parameters are well pre-trained and thus frozen.
The learnable modules include CM Block and the Predictor, which are multi-layer perceptrons. To prompt the model to learn to distinguish the target trials from other trials, we use the cross-entropy loss on $s_{cm}$ output by CM Block and $\hat{y}_{t}$ output by the Predictor respectively.

\section{Experiments}
\label{sec:exp}

\subsection{Experimental setup}
In the SASV Challenge 2022 \cite{jung2022sasv}, participants are restricted to utilise ASVspoof 2019 \cite{wang2020asvspoof} and VoxCeleb2 \cite{chung2018voxceleb2} datasets for training the anti-spoofing and ASV model, respectively. 

Three EERs, namely SV-EER, SPF-EER and SASV-EER are measured as the evaluation metrics, and SASV-EER is the main metric in the Challenge.

For the ASV models, we use Resnet34 \cite{he2016deep}, ECAPA-TDNN \cite{desplanques2020ecapa} and MFA-Conformer \cite{zhang2022conformer}.
For the countermeasure models, we use AASIST \cite{jung2021aasist}, AASIST-L, and RawGAT-ST \cite{tak2021end}, where AASIST-L is a light version of AASIST.
The fusion model in Figure~\ref{fig:method} is trained by Adam optimizer with an initial learning rate of 0.0001.
We set the batch size as 32, and the epoch number as 40.

\begin{table*}[ht]
\centering
\renewcommand\arraystretch{1.0}
\setlength\tabcolsep{12pt}
\caption{Performance of all systems on the ASVspoof 2019 development and evaluation sets.}
\label{tab:all-eer}
\begin{tabular}{lccccccc}
\hline
\hline
 & \multirow{2}{*}{\textbf{System}} &  \multicolumn{2}{c}{\textbf{SV-EER}} & \multicolumn{2}{c}{\textbf{SPF-EER}} & \multicolumn{2}{c}{\textbf{SASV-EER}} \\ 
  &  & Dev & Eval & Dev & Eval & Dev  & Eval \\ 
\hline
(A) & ECAPA-TDNN   & 1.64   & 1.86  & 20.28  & 30.75 & 17.37 & 23.84     \\
\hline
(B) & AASIST   & 46.01   & 49.24  & 0.07  & 0.67 & 15.86 & 24.38
 \\
\hline
(C1) & Baseline1 \cite{jung2022sasv}   & 32.88   & 35.32  & 0.06 & 0.67 & 13.07 & 19.31
 \\
(C2) & Baseline2 \cite{jung2022sasv}  & 12.87   & 11.48  & 0.13  & 0.78 & 4.85 & 6.37
 \\ 
\hline
(D) & MFA-Conformer + AASIST   & 1.48   & 1.47  & 0.20  & 1.08 & 0.88 & 1.35
 \\
 \hline
(E1) & SV-ALL + AASIST   & 1.42  & 1.30  & 0.27  & 1.61 & 0.81 & 1.41
 \\
(E2) & SV-ALL + AASIST-L  & 1.42  & 1.33  & 0.47 & 3.99 & 0.88 & 2.95
 \\
(E3) & SV-ALL + RawGAT-ST & 1.82  & 1.64  & 0.40  & 0.82 & 1.28 & 1.39
 \\
 \hline
(F1) & MFA-Conformer + CM-ALL-CAT-256   & 1.91  & 1.66 & 0.20 & 0.64 & 1.01 & 1.30
 \\
(F2) & ECAPA-TDNN + CM-ALL-CAT-256  & 1.39   & 1.73 & 0.20  & 0.74 & 0.81 & 1.40
 \\
(F3) & Resnet34 + CM-ALL-CAT-256 & 1.28 & 1.12  & 0.26 & 1.43 & 0.74 & 1.32
 \\
 \hline
(G1) & SV-ALL + CM-ALL-CAT-256 & 1.27   & 1.20  & 0.20 & 1.15 & 0.81 & 1.17
 \\
(G2) & SV-ALL + CM-ALL-CAT-512 & 1.35 & 1.15 & 0.20 & 1.12 & 0.74 & 1.14
 \\
(G3) & SV-ALL + CM-ALL-CAT-768 & 1.34 & 1.12 & 0.20 & 0.99 & 0.81 & \textbf{1.08}
 \\
(G4) & SV-ALL + CM-ALL-CAT-1024 & 1.28 & 1.21 & 0.20 & 0.83 & 0.74 & \textbf{1.08}
 \\
(G5) & SV-ALL + CM-ALL-CAT-2048 & 1.35 & 1.10 & 0.23 & 1.41 & 0.74 & 1.31
 \\
 \hline

(H1) & SV-ALL + CM-ALL-TAP-768 & 1.35 & 0.99 & 0.2 & 1.10 & 0.61 & \textbf{1.02}
 \\
(H2) & SV-ALL + CM-ALL-TSP-768 & 1.21 & 1.12 & 0.20 & 1.47 & 0.74 & 1.31
 \\
 (H3) & SV-ALL + CM-ALL-SAP-768 & 1.15 & 1.04 & 0.17 & 0.93 & 0.54 & \textbf{0.99}
 \\
(H4) & SV-ALL + CM-ALL-ASP-768 & 1.18 & 1.37 & 0.15 & 1.58 & 0.67 & 1.51
 \\
(H5) & SV-ALL + CM-ALL-TAP-1024 & 1.28 & 1.15 & 0.13 & 0.56 & 0.61 & \textbf{0.97}
 \\
(H6) & SV-ALL + CM-ALL-TSP-1024 & 1.11 & 1.12 & 0.20 & 1.77 & 0.61 & 1.43
 \\
(H7) & SV-ALL + CM-ALL-SAP-1024 & 1.15 & 1.16 & 0.20 & 1.45 & 0.61 & 1.28
 \\
 (H8) & SV-ALL + CM-ALL-ASP-1024 & 1.51 & 1.68 & 0.40 & 1.01 & 1.08 & 1.49
 \\
 \hline
 (I1) & Top-5 Ensemble & 1.08 & 1.01 & 0.20 & 0.71 & 0.67 & \textbf{0.89}
 \\
\hline
\hline
\end{tabular}
\end{table*}

\subsection{Experimental results and analysis}
As the requirements by SASV Challenge 2022, we evaluated systems on ASVspoof 2019 development and evaluation sets and reported SA-EER, SPF-EER and SASV EER, shown in Table \ref{tab:all-eer}. 
A and B denote systems using pure ASV or CM. We omit other systems using only single ASV or CM due to space limitation, yet they are with the same trend.
C1-C2 denote two baselines provided by the Challenge organizer. 
D-H8 are variants based on the proposed fusion strategy. D denote the fusion systems using only one ASV model and one CM model, e.g., 
'ECAPA-TDNN + AASIST' denotes the fusion of ECAPA-TDNN as the ASV model, and AASIST as the CM model.
Other systems using only one ASV and CM models are with the same trend.
E1-E3 denote systems fusing all three ASV models with one CM model. Note that D-E3 ignore the first-level fusion.
F1-F3 denote systems fusing one ASV model and all three CM models.
G1-H8 represent systems incorporating all three ASV models and all three CM models, but with different first-level fusion strategies and different sizes of CM Block. For example, 'SV-ALL + CM-ALL-CAT-768' denotes its first-level fusion uses concatenation (abbreviated as 'CAT' in the table), and fused $h_{cm}$ is projected to 768 dimensions in CM Block's first layer. I1 is the ensemble system involving the top-5 best evaluation set SASV-EER systems in A-H8. Figure \ref{fig:method} illustrates the histogram plots of three typical systems: (a) ECAPA-TDNN, a SOTA ASV system; (b) AASIST, a SOTA CM system; (c) Proposed top-5 ensemble system.

\subsubsection{Single-objective systems}
\textit{Only using speaker verification models.} ECAPA-TDNN performed well on the speaker verification sub-task and achieved 1.86\%, 1.38\% and 1.08\% SV-EERs on the evaluation set, respectively. However, they perform unacceptably on the anti-spoofing sub-task, yielding 30.75\%, 30.22\%, 29.76\% SPF-EER. Spoiled by the spoofing attacks, it is unpractical to perform SASV tasks using only the ASV models. 
Take ECAPA-TDNN model as an example, as shown in Figure \ref{fig:hist} (a), the pure ASV model can separate target and non-target trials while can hardly distinguish spoofing trials from genuine trials. This phenomenon is predictable because the objective of speaker verification models tends to erase the in-congruent device and environment information to more robustly identify speakers. However, these real-world traces should have helped to defend against spoofing attacks.

\textit{Only using anti-spoofing models.} In contrast, AASIST, the state-of-the-art CM model, can significantly discriminate spoofing utterances, but randomly guess on the speaker identification sub-task, mainly because of their speaker-unrelated objectives. They achieve SPF-EERs of 0.67\%, 0.84\%, 0.96\% on anti-spoofing, while yielding SV-EERs close to 50\% for speaker verification. This phenomenon can also be observed in Figure \ref{fig:hist} (b), where the target and non-target trials' distribution are almost totally overlapped. 

\subsubsection{Baseline systems}
Compared to the above single-objective models, two baseline systems reveal superiority in the SASV challenge. Among them, Baseline2 achieves a SASV-EER of 8.75\% on the evaluation set, which is better than Baseline1's 19.15\%. We argue that the reason is that the simple score-level fusion used in Baseline1 does not guarantee that ASV and CM scores belong to a unified space with a consistent magnitude. As shown in Figure \ref{fig:hist}, the ASV score ranges from -1 to 1, while the CM score ranges from -20 to 15. Straightforward addition will make the ASV score submerged by the CM score. In comparison, Baseline2 uses a trainable deep neural network to digest ASV and CM embeddings better, which can contribute to distinguishing the target from non-target and spoofing trials. However, the performance of Baseline2 on the speaker verification sub-task is still not satisfactory, which motivates us to propose the multi-model 
and multi-level fusion strategy.

\subsubsection{Proposed fusion systems}
The proposed strategy fuses three SOTA ASV models and three SOTA CM models. 
With different settings, variety of variant fusion systems (D-H8) are elaborately designed, and an additional ensemble system (I1) is constructed by integrating the top-5 best evaluation set SASV-EER systems from A-H8. Table \ref{tab:all-eer} illustrates the performance of all systems, and G1 is the best result of our previous work \cite{wu2022sasv}.

From Table \ref{tab:all-eer}, all the proposed systems outperform the baseline models with a large margin on the SASV-EER metrics, while retaining universally good performances on SV-EER and SPF-EER. 
The proposed models perform consistently well on the speaker verification task, and quite a few can reach or even surpass the performance of the SOTA ASV models. G1-H8 show better performance than D-F3 in general, benefiting from the more comprehensive model fusion. 

Using concatenation for the first-level fusion, we attempt different CM Block sizes. From G1 to G5, as we increase the size of the CM Block, the SASV-EER of the system decreases. 
Until adding a layer with an output dimension of 2048 to the bottom of the CM Block, the performance saturates.

Since G3 and G4 have the best SASV-EER, we employ their CM Block settings and further explore the impact of replacing concatenation with different pooling methods.
In H1-H8, under both CM block settings, TAP bring more benefits to the system than other methods. 
Both statistics pooling methods (TSP, ASP) are less effective than TAP, SAP, and concatenation. 
A possible reason is that we only compute the standard deviation for up to three CM embeddings, which is not stable during training. 
Rather than representing useful knowledge, the standard deviation seems to be more of a noise. 
We argue that if more CM models enroll, the system will benefit from statistics pooling methods, and we will leave it as future work.

The best individual system is with SASV-EER as 0.97\%. Moreover, the top-5 ensemble system (I1) achieves a SASV-EER as low as 0.89\%, which is a 86\% relative improvement compared to Baseline2. 
Figure \ref{fig:hist} (c) shows that the distributions of the target, non-target and spoofing trials are well-separated, which verifies the effectiveness of the proposed method.

\section{Conclusion}

This work proposes a novel multi-model and multi-level fusion strategy to tackle the SASV task.
The two-level fusion method can take advantage of the state-of-the-art ASV and CM models.
As a result, the best single fusion system achieves the SASV-EER of 0.97\%.
What's more, by ensembling the top-5 systems, the final SASV-EER reaches 0.89\% on the evaluation set, which is 86\% relative reduction compared to the best baseline, Baseline2, and 24\% relative reduction compared to our previous work \cite{wu2022sasv}.
In the future work, we will introduce more CM models to investigate the potential of the proposed method.


\bibliographystyle{IEEEtran}


\end{document}